\documentclass[11pt,a4paper,preprintnumbers,notitlepage,nofootinbib]{revtex4-1}

\usepackage[T1]{fontenc} 
\usepackage{graphicx}
\usepackage{xspace}
\usepackage{amsmath,amssymb,mathtools}
\RequirePackage{flushend}
\RequirePackage[colorlinks,citecolor=blue,urlcolor=blue,linkcolor=blue]{hyperref}
\usepackage[utf8]{inputenc}
\usepackage{color}
\usepackage{natbib}
\usepackage{enumitem}

\def\nn{\nonumber}
\newcommand{\veff}{V_{\text{eff}}}
\newcommand{\vtree}{V^{(0)}}
\newcommand{\vone}{V^{(1)}}
\newcommand{\vtwo}{V^{(2)}}

\DeclareMathOperator{\tr}{\text{tr}}

\def\msbar{{\ensuremath{\overline{\rm MS}}}\xspace}

\def\tri{{\ensuremath{\lambda_{hhh}}}\xspace}

\begin{document}

\preprint{DESY-21-059}
\preprint{OU-HET-1098}

\title{Two-loop corrections to the Higgs trilinear coupling in BSM models with classical scale invariance \\\textit{{\mdseries{\small Talk presented at the International Workshop on Future Linear Colliders (LCWS2021), 15-18 March 2021. C21-03-15.1.}}}}

\author{Johannes~Braathen\footnote{Speaker.}} 
\email{johannes.braathen@desy.de}
\affiliation{Deutsches Elektronen-Synchrotron DESY, Notkestra\ss{}e 85, D-22607 Hamburg, Germany.}

\author{Shinya~Kanemura} 
\email{kanemu@het.phys.sci.osaka-u.ac.jp}
\affiliation{Department of Physics, Osaka University, Toyonaka, Osaka 560-0043, Japan.}

\author{Makoto Shimoda}
\email{m_shimoda@het.phys.sci.osaka-u.ac.jp}
\affiliation{Department of Physics, Osaka University, Toyonaka, Osaka 560-0043, Japan.}

\begin{abstract}

Classical scale invariance (CSI) is an attractive concept for BSM model building, explaining the apparent alignment of the Higgs sector and potentially relating to the hierarchy problem. Furthermore, a particularly interesting feature is that the Higgs trilinear coupling \tri is \emph{universally predicted at one loop in CSI models}, and deviates by 67\% from its (tree-level) SM prediction. 
This result is however modified at two loops, and we review in these proceedings our calculation~\cite{Braathen:2020vwo} of leading two-loop corrections to \tri in models with classical scale invariance, taking as an example a CSI scenario of a Two-Higgs-Doublet Model. We find that the inclusion of two-loop effects allows distinguishing different scenarios with CSI, although the requirement of reproducing the known 125-GeV Higgs-boson mass severely restricts the allowed values of \tri.

\end{abstract}

\maketitle

\section{Introduction}

The Higgs trilinear coupling \tri is a unique probe to investigate the structure of the Higgs sector and the nature of the electroweak phase transition (EWPT), as well as to search for indirect signs of New Physics. While the discovery of Higgs boson has established the mechanism of electroweak symmetry breaking (EWSB), little is known about the structure of the Higgs potential. Currently, only the location of the electroweak (EW) minimum -- obtained from the Higgs vacuum expectation value (VEV) -- and the curvature of the potential around this minimum -- related to the Higgs mass -- are known. The determination of \tri is therefore of great importance, as this coupling decides behaviour of the Higgs potential away from its EW minimum. It is for example known that a large deviation (of order 20\%) of \tri from its SM prediction is necessary to have an EWPT of strong first-order nature~\cite{Grojean:2004xa,*Kanemura:2004ch} -- which is itself a requirement for successful EW baryogenesis scenarios~\cite{Sakharov:1967dj,*Kuzmin:1985mm,*Cohen:1993nk}.

The Higgs trilinear coupling can also play an important role in scrutinising possible Beyond-the-Standard-Model (BSM) theories. One such example comes from theories with classical scale invariance (CSI), in which mass-dimensionful terms are forbidden in the Lagrangian at the tree level. While this symmetry is often imposed at the Planck scale to address the hierarchy problem, requiring CSI at (or around) the EW scale is also of great interest for model building, for a number of phenomenological reasons. First, such models constitute an ideal example of \emph{alignment}~\cite{Gunion:2002zf} \emph{without decoupling}: the couplings of the Higgs boson -- which corresponds to the direction in field space along which the EW symmetry is broken radiatively, \`{a} la Coleman-Weinberg ~\cite{Coleman:1973jx,*Gildener:1976ih} -- are automatically SM-like at the tree level, and at the same time, BSM states cannot be decoupled because of the absence of mass terms in the Lagrangian. Another discriminative feature of these models is that at one loop they \emph{universally} predict the value of \tri to deviate by 67\% from the corresponding tree-level SM prediction, as found in Ref.~\cite{Hashino:2015nxa}. 

Although this CSI prediction deviates significantly from the SM, it is still well within what is presently allowed by experimental results. Indeed, the current best limits on \tri have been obtained by the ATLAS collaboration using LHC Run 2 single-Higgs production searches, and give $-3.2 < \tri/\lambda_{hhh}^\text{SM} < 11.9$ at 95\% confidence level (CL)~\cite{ATL-PHYS-PUB-2019-009}. However, the determination of \tri will be substantially improved at future colliders: the HL-LHC could reach an accuracy of 50\%~\cite{Cepeda:2019klc}, while a lepton collider, such as the ILC, might achieve a precision of some tens of percent~\cite{Fujii:2017vwa,*Roloff:2019crr} (with high centre-of-mass energies) -- for a complete review, see $e.g.$ Ref.~\cite{deBlas:2019rxi}. The large one-loop deviation in \tri predicted in CSI models would then be detectable in a foreseeable future. 

It is then also important to investigate how higher-order effects affect this prediction, especially as two-loop corrections to \tri have been found to be significant in several non-CSI models~\cite{Braathen:2019pxr,*Braathen:2019zoh} (see also Ref.~\cite{Senaha:2018xek}). For this reason, we computed in Ref.~\cite{Braathen:2020vwo} the dominant two-loop corrections to \tri in models with CSI. Crucially, we found that the inclusion of two-loop contributions spoils the universality of the prediction for \tri ~-- a result we illustrated both in an $N$-scalar model and in a CSI variant of Two-Higgs-Doublet Model (2HDM). Even though the allowed ranges of BSM parameters are strongly constrained in CSI models, in particular by perturbative unitarity~\cite{Lee:1977eg} and the requirement of reproducing the known 125-GeV mass of the SM-like Higgs boson, the two-loop corrections to \tri give rise to a further deviation of 10-30\% from the SM, and can allow distinguishing different scenarios of a given CSI model. Summarising our findings from Ref.~\cite{Braathen:2020vwo}, we discuss in these proceedings how the calculation of the Higgs trilinear coupling in CSI theories is modified at two loops, before showing examples of numerical results for a CSI 2HDM. 

\section{The Higgs trilinear coupling in CSI theories}
We begin by describing the effective-potential calculation of the Higgs trilinear coupling in CSI theories, first reproducing the known universal one-loop result of $e.g.$ Ref.~\cite{Hashino:2015nxa}, and next illustrating how this is altered at two loops. At both one- and two-loop orders, the computation of contributions to the effective potential $\veff$ is greatly simplified in CSI theories. This is because all mass-dimensionful parameters in the Lagrangian vanish due to the requirement of classical scale invariance, and thus the only possible source of mass is through a particle's coupling to the Higgs boson and its VEV $v$. In turn, this means that the field-dependent mass $m_i(h)$ of any particle $i$ can always be expressed in terms of its mass-independent mass $m_i$ as
\begin{align}
\label{EQ:fielddepmass}
m_i^2(h)=m_i^2\left(1+\frac{h}{v}\right)^2\,,
\end{align}
no matter the nature of field $i$. 

\subsection{\tri at one loop}

At one loop, the effective potential is calculated using the well-known supertrace formula~\cite{Jackiw:1974cv}. Together with eq.~(\ref{EQ:fielddepmass}), one finds that along the Higgs-field direction $\veff$ can be written as
\begin{align}
 \veff(h)=A^{(1)}\ (v+h)^4+B^{(1)}\ (v+h)^4\log\frac{(v+h)^2}{Q^2}\,,
\end{align}
where $Q$ is the renormalisation scale and the dimensionless coefficients $A^{(1)}$ and $B^{(1)}$ read 
\begin{align}
 A^{(1)}&\equiv \frac{1}{64\pi^2v^4}\bigg\{\tr\left[M_S^4\left(\log\frac{M_S^2}{v^2}-\frac32\right)\right]-4\tr\left[M_f^4\left(\log\frac{M_f^2}{v^2}-\frac32\right)\right]+3\tr\left[M_V^4\left(\log\frac{M_V^2}{v^2}-\frac56\right)\right]\bigg\}\,,\nn\\
 B^{(1)}&\equiv \frac{1}{64\pi^2v^4}\big\{\tr\left[M_S^4\right]-4\tr\left[M_f^4\right]+3\tr\left[M_V^4\right]\big\}\,,
\end{align}
with $M_S,\, M_f,\, M_V$ the scalar, fermion, and gauge-boson mass matrices respectively. 

If we want to derive \tri by an effective-potential calculation, we must also take into account the conditions imposed by the lower-order derivatives of $\veff$, $i.e.$ the tadpole equation and the Higgs effective-potential mass. Specifically, these yield the following relations
\begin{align}
 \frac{\partial \veff}{\partial h}\bigg|_{h=0}=0=&\ 2v^3\left(2A^{(1)}+B^{(1)}+2B^{(1)}\log\frac{v^2}{Q^2}\right)\,,\nn\\
 \frac{\partial^2 \veff}{\partial h^2}\bigg|_{h=0}\equiv [M_h^2]_{\veff}=&\ 2v^2\left(6A^{(1)}+7B^{(1)}+6B^{(1)}\log\frac{v^2}{Q^2}\right)=8v^2B^{(1)}\,,
\end{align}
which allow determining $A^{(1)}$ and $B^{(1)}$ entirely in terms of the Higgs VEV $v$ and its effective-potential (or curvature) mass $[M_h^2]_{\veff}$. It then follows that the third derivative of $\vone$ is also entirely fixed in terms of $v$ and $[M_h^2]_{\veff}$ and one obtains~\cite{Hashino:2015nxa}
\begin{align}
\label{EQ:oneloop_lhhh}
 \lambda_{hhh}\equiv\frac{\partial^3\veff}{\partial h^3}\bigg|_{h=0}=\frac{5[M_h^2]_{\veff}}{v}=\frac53(\lambda_{hhh}^\text{SM})^{(0)}\,,
\end{align}
where $(\lambda_{hhh}^\text{SM})^{(0)}$ denotes the tree-level prediction for the trilinear coupling in the SM. It is important to stress here that this result is \emph{totally independent} of the particle content or scenario of CSI model.

\subsection{\tri at two loops}

This discussion is however modified if one considers the inclusion of two-loop corrections to the effective potential, and to the Higgs trilinear coupling derived from it. Indeed, two-loop contributions to $\veff$ consist of one-particle-irreducible vacuum bubble diagrams, and their analytical expressions are known to contain not only non-logarithmic and logarithmic terms, but also squared-logarithmic terms -- see $e.g.$ Ref.~\cite{Ford:1992pn}. This means that the potential is now of the form
\begin{align}
\label{EQ:veff_2l}
 \veff=A\ (v+h)^4+B\ (v+h)^4\log\frac{(v+h)^4}{Q^2}+C\ \log^2\frac{(v+h)^2}{Q^2}\,,
\end{align}
where $A$ and $B$ receive both one- and two-loop contributions (respectively $A^{(1)},\ B^{(1)}$ and $A^{(2)},\ B^{(2)}$), while $C$ is a new type of coefficient that only enters from two loops. 

If one then repeats the steps described above for the calculation of \tri, $A$ and $B$ can still be eliminated using the tadpole equation and Higgs curvature mass relation, however, $C$ remains. One obtains finally at two loops \vspace{-0.5cm}
\begin{align}
\label{EQ:lhhh_2l}
\tri=\frac{5[M_h^2]_{\veff}}{v}+32 Cv\,,
\end{align}
$i.e.$ there is a new contribution to the Higgs trilinear coupling that depends on the squared-logarithmic piece of the two-loop effective potential. Moreover, because the effective-potential contributions are typically model-dependent, this result means that the universality of the result for \tri is \emph{lost} at two loops.

\section{$\lambda_{hhh}$ at two loops in a CSI 2HDM}
We consider now a specific model, namely the classically scale-invariant version of the Two-Higgs-Doublet Model, devised in Ref~\cite{Lee:2012jn} (and studied also in Refs.~\cite{Lane:2018ycs,Lane:2019dbc,Brooijmans:2020yij}). Although it shares many similarities to the usual 2HDM (see $e.g.$ Ref.~\cite{Branco:2011iw} for a review), the CSI 2HDM differs by some key aspects which we briefly recall here. First of all, the scalar potential contains no mass terms, only quartic interactions, and is written, assuming CP conservation, in terms of two $SU(2)_L$ doublets $\Phi_1$ and $\Phi_2$ of hypercharge $\frac12$ as
\begin{align}
 \vtree=\frac12\lambda_1|\Phi_1|^4+\frac12\lambda_2|\Phi_2|^4+\lambda_3|\Phi_1|^2|\Phi_2|^2+\lambda_4|\Phi_1^\dagger\Phi_2|^2+\frac12\lambda_5\big[(\Phi_1^\dagger\Phi_2)^2+\text{h.c.}\big]\,.
\end{align}
The physical spectrum is the same as in the usual 2HDM, with two CP-even Higgs bosons $h$ and $H$ (we assume the lighter one $h$ to be SM-like), a CP-odd Higgs boson $A$, and a charged Higgs boson $H^\pm$. However, a particular feature of the CSI 2HDM is that its scalar sector is automatically aligned at tree level -- this can be understood as a consequence of the relations between quartic couplings arising from the tadpole equations (in the absence of mass terms in $\vtree$). 

The steps of our computation of the leading two-loop contributions to \tri can be summarised as follows (for more details we refer the reader to Ref.~\cite{Braathen:2020vwo}): \vspace{-.3cm}
\begin{enumerate}
 \item We start by computing the dominant contributions to the two-loop effective potential $\vtwo$, employing expressions from Refs.~\cite{Martin:2001vx,*Martin:2003qz,*Degrassi:2009yq,*Braathen:2016cqe}. In the CSI 2HDM, these are diagrams involving the heavy BSM scalars and the top quark.\vspace{-.3cm}
 \item We then extract from these the coefficient of the terms proportional to $\log^2 (v+h)^2/Q^2$ -- what we had defined as $C$ in eq.~(\ref{EQ:veff_2l}).\vspace{-.3cm}
 \item This allows us to derive directly an \msbar scheme expression for \tri, using eq.~(\ref{EQ:lhhh_2l}).\vspace{-.3cm}
 \item Finally, to obtain an on-shell scheme result for the Higgs trilinear coupling, we include the necessary finite counterterms to express our result in terms of OS renormalised quantities, and we include finite wave-function renormalisation effects.\vspace{-.3cm}
\end{enumerate}
Detailed expressions for the two-loop corrections to \tri in the CSI 2HDM can be found in section IV and appendix C of Ref.~\cite{Braathen:2020vwo}. 
At this point, we should emphasise that we are performing here an effective-potential calculation of \tri, and thereby we are not including effects from non-vanishing external momenta -- however, we expect these to be subleading. We will also neglect any loop induced deviation from the alignment occurring at tree level -- these effects are known to be small, $c.f.$ Refs.~\cite{Lane:2019dbc,Braathen:2017izn}.

\begin{figure}[t]
 \includegraphics[width=.7\textwidth]{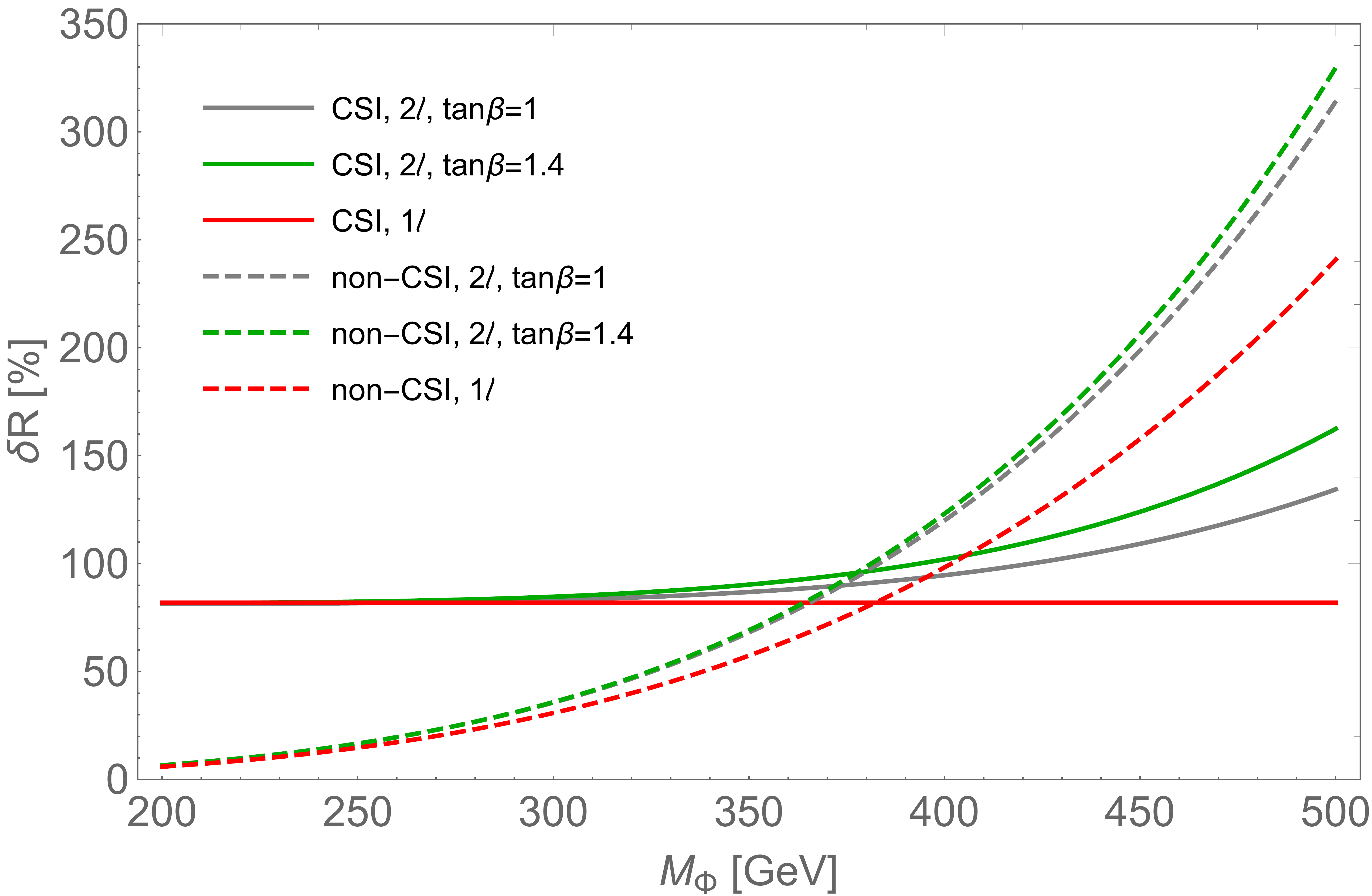}
 \caption{Comparison of the BSM deviation $\delta R$ computed in scenarios of the 2HDM, with (solid curves) and without (dashed curves) CSI, as a function of the degenerate pole mass $M_\Phi$ of the BSM scalars. Red curves present one-loop values, while grey and green curves are two-loop results for $\tan\beta=1$ and $\tan\beta=1.4$ respectively.}
 \label{FIG:2HDM_compareCSInonCSI}
\end{figure}

In the following, we illustrate our calculation with numerical examples, in which we will always show the BSM deviation $\delta R$ in the Higgs trilinear coupling, defined as $\delta R\equiv\lambda_{hhh}^\text{2HDM}/\lambda_{hhh}^\text{SM}-1$.
In figure~\ref{FIG:2HDM_compareCSInonCSI}, we present results for $\delta R$ as a function of the degenerate mass of the BSM scalars $M_\Phi=M_H=M_A=M_{H^\pm}$, in the CSI 2HDM (solid curves) and in a non-CSI 2HDM (dashed curves) -- for the latter we employ expressions from Refs.~\cite{Braathen:2019pxr,*Braathen:2019zoh}. The red curves correspond to the BSM deviations obtained at one loop, while the gray and green ones show the two-loop values of $\delta R$ with $\tan\beta=1$ and $\tan\beta=1.4$ respectively. 
Concentrating first on the CSI results, we can immediately observe that in contrast to the constant one-loop result\footnote{Note that this value differs from the 67\% mentioned in the previous section, because -- per the definition of $\delta R$ -- we compare here the one-loop CSI result for \tri to the \emph{one-loop} (rather than tree-level) SM prediction.} of $(\delta R)^{(1)}\simeq 82\%$, at two loops $\delta R$ depends on both $M_\Phi$ and $\tan\beta$. In other words, the inclusion of two-loop corrections to \tri allows distinguishing different parameter points of the CSI 2HDM. It is also interesting to note, that the new two-loop contributions always yield an increase of $\delta R$, thereby making the Higgs trilinear coupling more easily accessible at experiments. 
Next, comparing the solid and dashed curves, we find that the behaviour of \tri changes drastically between the variants of 2HDM with or without CSI. This is mostly due to the large difference at one loop between on the one hand, a large constant deviation in the CSI case, and one the other hand, a correction proportional to $M_\Phi^4$ in the non-CSI case. Due to this, for small values of the BSM scalar masses, larger BSM effects are found in the CSI scenario, while for heavy BSM scalars the largest deviations occur in the non-CSI scenario, as both one- and two-loop contributions grow rapidly with $M_\Phi$. 

\begin{figure}[t]
 \includegraphics[width=.9\textwidth]{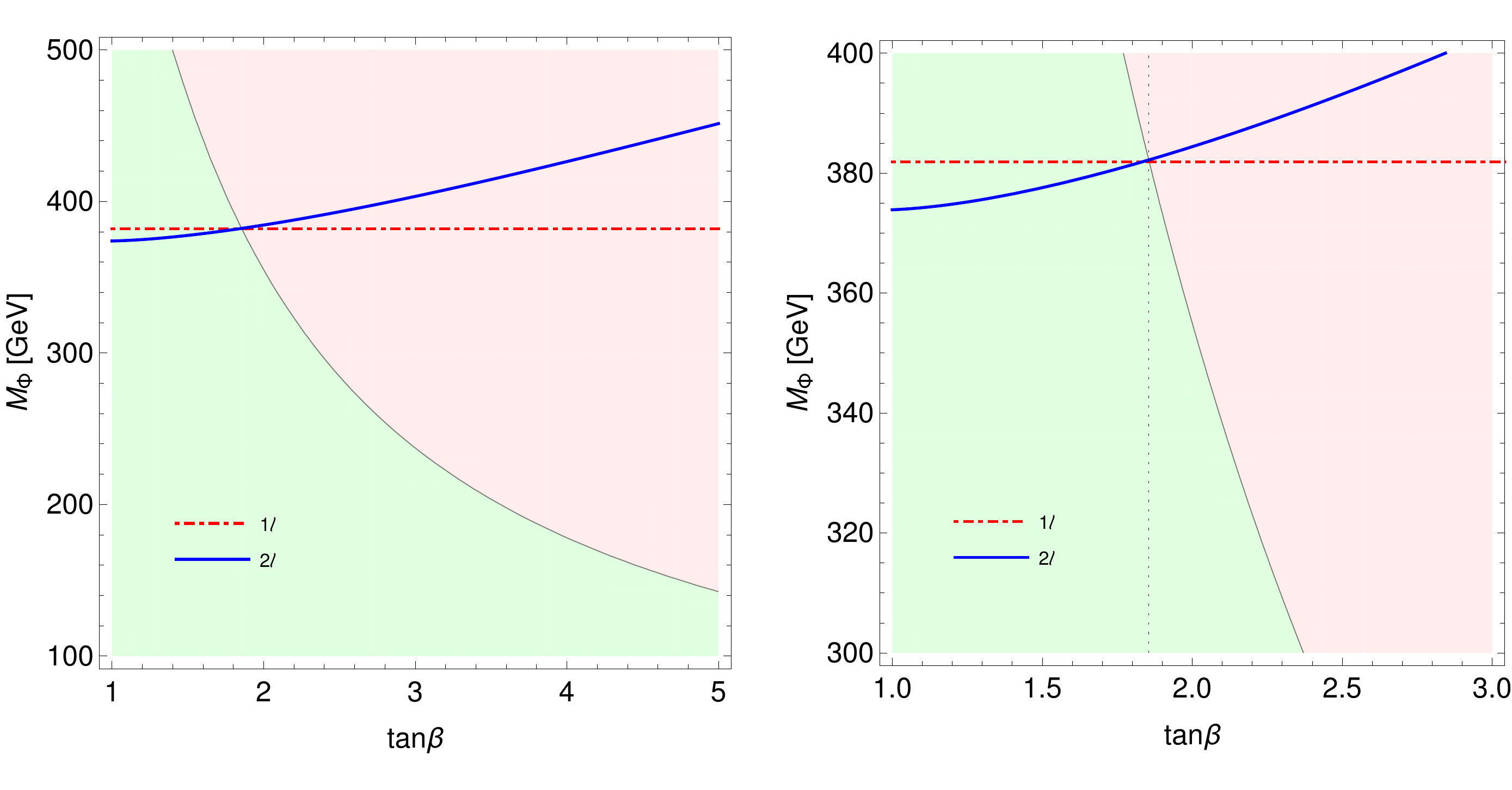}\
 \caption{Regions of the $\tan\beta$ and $M_\Phi$ parameter space of the CSI 2HDM allowed (light green) and excluded (light red) by the requirement of tree-level perturbative unitarity~\cite{Kanemura:1993hm}. Additionally, the red and blue curves give the values of $M_\Phi$ computed as a function of $\tan\beta$, at one and two loops respectively, from the requirement of correctly reproducing $M_h=125\text{ GeV}$. The right-hand side plot is a zoom of the left one. }
 \label{FIG:CSI2HDM_unitaritycontours}
\end{figure}

When making figure~\ref{FIG:2HDM_compareCSInonCSI}, we have verified that the EW vacuum remains the true minimum of $\veff$ and that perturbative unitarity~\cite{Lee:1977eg} is fulfilled for the entire range of $M_\Phi$ and values of $\tan\beta$ we considered. In figure~\ref{FIG:CSI2HDM_unitaritycontours}, we show in light green and light red the regions of the $\{\tan\beta,\,M_\Phi\}$ parameter plane that are respectively allowed and excluded from the constraints of tree-level perturbative unitarity, following the results of Ref.~\cite{Kanemura:1993hm} (see also Ref.~\cite{Akeroyd:2000wc}). Another important constraint on the BSM parameters is given by the SM-like Higgs boson mass: as the Higgs boson corresponds to the flat direction of the tree-level potential, its mass is generated entirely at loop level. This yields a relation between the Higgs pole mass, the (known) SM input parameters, and the BSM parameters of the CSI 2HDM -- $\tan\beta$ and the BSM scalar masses -- which can be used to extract the value of one of the BSM parameters. The red and blue lines in figure~\ref{FIG:CSI2HDM_unitaritycontours} give the values of $M_\Phi$ obtained as a function of $\tan\beta$ from this condition -- the expression for this relation can be found in eq.~(IV.21) of Ref.~\cite{Braathen:2020vwo} (see also eq.~(C.23) for the case of non-degenerate BSM scalar masses). At the one-loop level, there is no dependence on $\tan\beta$, because $M_h$ is independent of $\tan\beta$ at this order, however, this is modified at two loops and one can derive $M_\Phi=M_\Phi(\tan\beta)$ -- or equivalently $\tan\beta=\tan\beta(M_\Phi)$. We can also notice that only a limited range of $\tan\beta$ and $M_\Phi$ are allowed when taking into account both perturbative unitarity and the Higgs mass condition.

\begin{figure}[t]
 \centering
 \includegraphics[width=.7\textwidth]{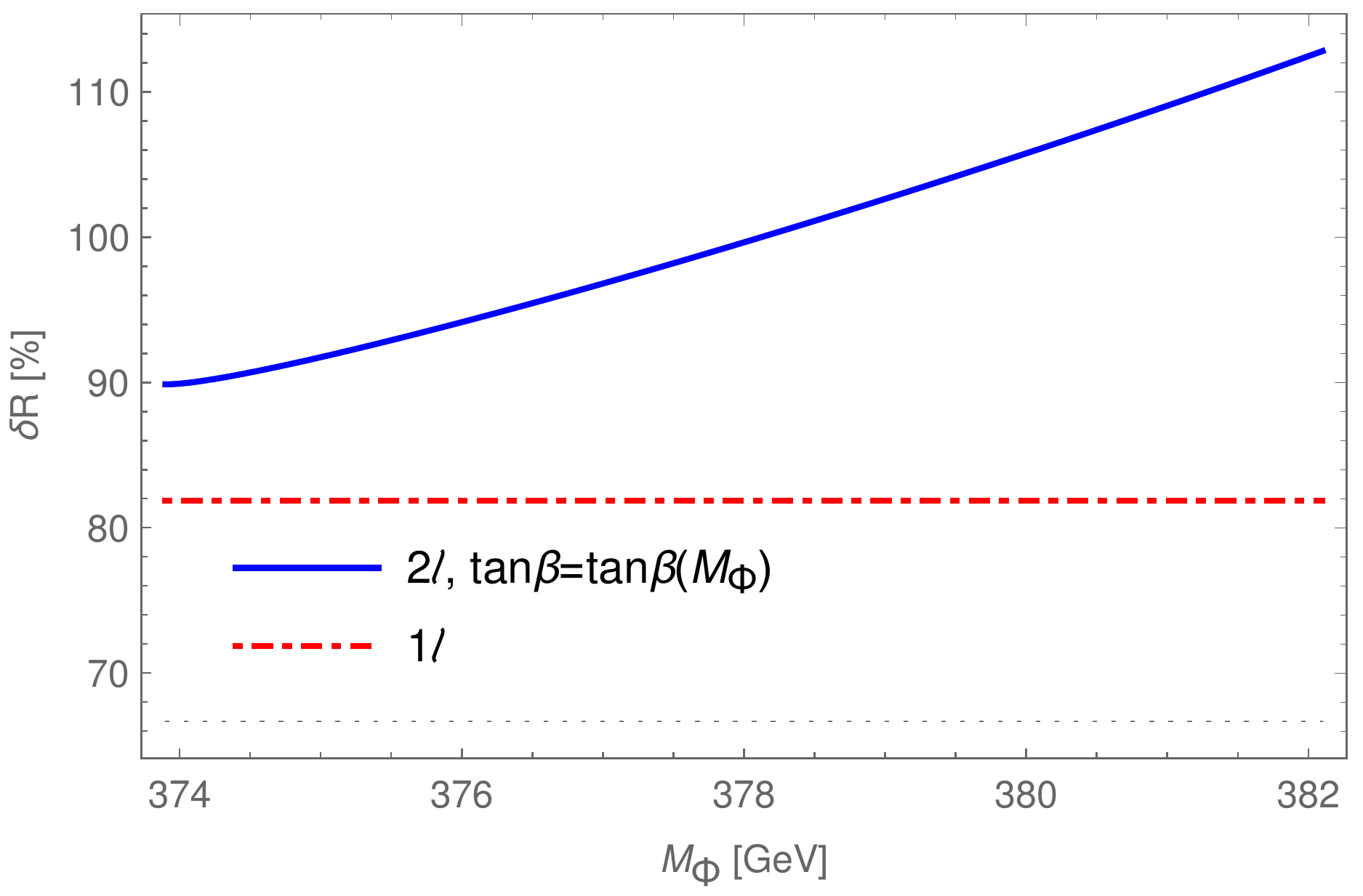}
 \caption{BSM deviation $\delta R$ of the Higgs trilinear coupling $\lambda_{hhh}$ computed at two loops in the CSI 2HDM with respect to its SM prediction as a function of the degenerate mass of the BSM scalars $M_\Phi=M_H=M_A=M_{H^\pm}$. $\tan\beta$ is computed as a function of $M_\Phi$ in order to reproduce $M_h=125\text{ GeV}$ and the range of $M_\Phi$ corresponds to the values allowed under perturbative unitarity ($c.f.$ fig.~\ref{FIG:CSI2HDM_unitaritycontours}). The red dot-dashed and black dotted lines show the comparison of the one-loop CSI value of $\lambda_{hhh}$ with respectively the one-loop effective-potential and tree-level results in the SM.} 
 \label{FIG:CSI2HDM_deltaRnonCSIconstanbeta}
\end{figure}

If we finally use this condition from the Higgs mass to constraint $\tan\beta$ as a function of $M_\Phi$ in the computation of two-loop corrections to \tri, we obtain the blue curve in figure~\ref{FIG:CSI2HDM_deltaRnonCSIconstanbeta}, which shows the two-loop BSM deviation $\delta R$ in the narrow allowed range of the degenerate BSM scalar mass $M_\Phi$. We then find that the BSM deviation at two loops could range from 90\% to 112\% -- in other words a further positive deviation of approximately $10-30\%$ in addition to the one-loop result. Comparing figures~\ref{FIG:2HDM_compareCSInonCSI} and~\ref{FIG:CSI2HDM_deltaRnonCSIconstanbeta}, we can observe that the need to reproduce $M_h=125\text{ GeV}$ severely limits the possible values of \tri, although the two-loop contributions remain significant with respect to their one-loop counterparts. 
Returning lastly to fig.~\ref{FIG:2HDM_compareCSInonCSI}, we can observe that this range of values of $M_\Phi$ and of $\delta R$ corresponds to where the BSM deviations in the CSI and non-CSI version of the 2HDM overlap. In Ref.~\cite{Braathen:2020vwo}, we also obtained a similar result for an $N$-scalar model. It therefore seems that it may difficult to use the value of \tri alone to distinguish scenario of a BSM theory with or without classical scale invariance. Using \texttt{HiggsBounds}~\cite{Bechtle:2008jh,*Bechtle:2011sb,*Bechtle:2013wla,*Bechtle:2015pma,*Bechtle:2020pkv} (for which the necessary inputs were computed using a \texttt{SPheno}~\cite{Porod:2003um,*Porod:2011nf} based spectrum generator for the CSI 2HDM, created with \texttt{SARAH}~\cite{Staub:2008uz,*Staub:2009bi,*Staub:2010jh,*Staub:2012pb,*Staub:2013tta}) we verified that the parameter points in fig.~\ref{FIG:CSI2HDM_deltaRnonCSIconstanbeta} were not excluded by experimental searches.

\vspace{-.5cm}

\section{Conclusion}
We have summarised here our recent paper~\cite{Braathen:2020vwo}, in which we performed the first explicit two-loop calculation of the Higgs trilinear coupling \tri in theories with CSI. Our results are significant in that they match the level of accuracy reached for non-CSI extensions of the SM in Refs.~\cite{Braathen:2019pxr,*Braathen:2019zoh}. Furthermore, we have shown that the inclusion of two-loop contributions breaks the universality of the prediction of \tri in CSI models. This implies that accessing the value \tri could allow distinguishing different scenarios and different parameter points of a given CSI model. However, we have also pointed out the important fact that perturbative unitarity, as well as the requirement of generating the correct SM-like Higgs mass at loop level, strongly constrain the allowed range of parameters in CSI theories, and in turn the possible values of \tri. For the CSI 2HDM considered here, we predict a BSM deviation of 90-110\% of the Higgs trilinear coupling from its (two-loop) SM prediction. Finally, while it may be difficult to use a measurement of the Higgs trilinear coupling on its own to distinguish variants of a BSM theory with or without CSI -- as we saw in figs.~\ref{FIG:2HDM_compareCSInonCSI} and \ref{FIG:CSI2HDM_deltaRnonCSIconstanbeta} -- we note that this could be achieved using the \emph{synergy} of a measurement of \tri with either signals from collider searches, or gravitational-wave experiments -- as studied for instance in Ref.~\cite{Hashino:2016rvx}.\vspace{-.5cm}

\subsection*{Acknowledgments}
This work is supported by JSPS, Grant-in-Aid for Scientific Research, No. 16H06492, 18F18022, 18F18321 and 20H00160. This work is also partly supported by the Deutsche Forschungsgemeinschaft (DFG, German Research Foundation) under Germany’s Excellence Strategy – EXC 2121 “Quantum Universe” – 390833306.

\bibliography{BKSproc}

\end{document}